\documentclass[showpacs,aps,graphicx,twocolumn]{revtex4}
\usepackage{graphicx}

\begin{document}

\title{Efficient multipartite entanglement  concentration of electron-spin state with charge detection}

\author{Lan Zhou,$^{1,2}$\footnote{Email address:
zhoul@njupt.edu.cn} Jiong Liu$^2$, Sheng-Yang Zhao$^2$ }
\address{$^1$College of Mathematics \& Physics, Nanjing University of Posts and Telecommunications, Nanjing,
210003, China\\
 $^2$Institute of Signal Processing  Transmission, Nanjing
University of Posts and Telecommunications, Nanjing, 210003,  China\\}

\begin{abstract}
We present two entanglement concentration protocols (ECPs) for
arbitrary three-electron W state based on their
charges and spins. Different from other ECPs, in both two ECPs, with the help of the electronic polarization beam splitter (PBS) and charge detection, the less-entangled W
state can be concentrated into a maximally entangled state only with some single charge qubits. The second ECP is more optimal than the first one, for by constructing the complete parity check gate, the second ECP can be used repeatedly to further concentrate the less-entangled state and obtain a higher success probability. Therefore, both the ECPs, especially the second one may be
useful in current quantum information processing.
\end{abstract}
 \maketitle

\section{Introduction}
Quantum entanglement plays an important role in current quantum
information processing and transmission
\cite{rmp}. The maximally entangled states
may be the most important resources in practical quantum
communication and computation tasks \cite{Ekert91,rmp,teleportation,
densecoding,QSTS,QKD,densecoding1,densecoding2,QSDC,QSS1,QSS2,QSS3}. In most of quantum communication
protocols, such as the quantum teleportation \cite{teleportation}, quantum key distribution (QKD) \cite{Ekert91,QKD}, quantum densecoding \cite{densecoding,densecoding1,densecoding2}, and quantum state sharing (QSS) \cite{QSS1,QSS2,QSS3}, they need the maximally entangled state to setup the
quantum channel. Unfortunately, the practical transmission channel
always contain noise, which will degrade the quality of the
entanglement, and make the maximally entangled state become a mixed
state or a less-entangled state \cite{Pan1}.

The method for distilling a mixed state into a maximally entangled
state is called the entanglement purification \cite{C.H.Bennett1,Pan1}. On the other hand, the way for
distilling a pure less-entangled state into a maximally entangled
state, which will be detailed
here, is called the entanglement concentration,
\cite{C.H.Bennett2,swapping1,swapping2,Yamamoto1,zhao1,shengpra2,shengqic,dengpra,shengpra3}.
The first entanglement concentration protocol (ECP) was proposed by Bennett \emph{et al.} in 1996, which
is called as the Schmidt projection method \cite{C.H.Bennett2}. In the protocol, they need
some collective and nondestructive measurements, which are not easy
to realize under current experimental condition.
In 1999, Bose \emph{et al}. showed that the entangle swapping could
also be used to achieve the task of ECP \cite{swapping1}. This
protocol was developed by Shi \emph{et al.} in 2000
\cite{swapping2}. In 2001, two similar ECPs based on the linear optical elements were
proposed by Yamamoto \emph{et al.} and Zhao \emph{et al.},
respectively \cite{Yamamoto1,zhao1}. The basic idea of these two ECPs is that they adopt the optical polarization beam splitter to complete the parity check measurement for photons. However, they require the sophisticated
single-photon detectors to make the photon number detection, which is not likely to be available. These
protocols were developed by Sheng \emph{et al.} in 2008
\cite{shengpra2}. They used the cross-Kerr nonlinearity to construct
the quantum nondemolition detector to act the roles of both parity
check measurement and single photon detector. In 2010, the ECP for
single-photon entanglement was also proposed \cite{shengqic}.

Up to now, most of the ECPs described above are focused on the bipartite entangled photon systems and there are only
several multipartite ECPs,
 In 2003, Cao and Yang
proposed an ECP for W state with joint unitary transformation
\cite{cao}. In 2007, Zhang \emph{et al.} proposed an ECP based on
the collective Bell-state measurement \cite{zhanglihua}. In 2010, Wang\emph{ et al.}  proposed an ECP for
a special  W state
$\alpha|HVV\rangle+\beta(|VHV\rangle+|VVH\rangle)$\cite{wanghf}, where $|H\rangle$ and $|V\rangle$
represent the horizontal and the vertical polarizations of photons,
respectively. Later, Yildiz proposed an ECP for three-qubit asymmetric W states \cite{yildiz}. In 2012, Sheng \emph{et al.} proposed an ECP for W state with the help of cross-Kerr nonlinearity \cite{shengwstate}.

For quantum communication,  photons are the best candidates for
carrying the information due to its fast transmission and easy
manipulation. Actually, the conduction electrons are also
good candidates in quantum information processing, especially in
quantum commutation. In 2004, by using charge detector, Beenakker
\emph{et al.} broke through the obstacle of the no-go theorem and
constructed a controlled-not (CNOT) gate with conduction electrons
\cite{Beenakker,nogo}. In their protocol, the charge detector can
distinguish the occupation of the charge number 1 from 0 and 2, but
it can not distinguish the charge number 0 and 2. Subsequently,
protocols of preparing cluster states \cite{zhangxl}, entanglement
purification and concentration\cite{feng,shengpla,shengpla2,dengelectron,wangcelec,wangchuanpra}, and other quantum information protocols
were proposed \cite{trauzettel,ionicioiu,eleccnot}.
As shown in Refs. \cite{hu1,hu2,hu3}, the
interaction between the polarizations of photons and the electron spins of quantum dots in optical microcavities can also
be used to perform the quantum information processing.

In this paper, we present two ECPs for three-electron less-entangled W state with the help of charge detection.
In the first ECP, we adopt one pair of less-entangled W state and two auxiliary electrons
to perform the concentration. With the help of the polarization beam splitter (PBS) and charge detection, the less-entangled W state can be distilled to the maximally entangled W state with some probability. In the second protocol, we construct the complete parity check gate and make the protocol can be used repeatedly to obtain a higher success probability.

This paper is organized as follows: In Sec. II, we explain the first
ECP with the help of the PBS and charge detection. In Sec.
III, we explain the second ECP with the complete parity
check gate. In Sec. IV, we make a discussion and summary.

\section{Efficient ECP with electronic PBS}
The schematic drawing of our ECP is shown in Fig. 1. Our ECP includes two steps. In the first concentration step, we suppose Alice, Bob, and Charlie share the three-electron
less-entangled W state from S$_{1}$, which can be described as
\begin{eqnarray}
|\Phi\rangle_{a1b1c1}&=&\alpha|\downarrow\rangle_{a1}|\uparrow\rangle_{b1}|\uparrow\rangle_{c1}
+\beta|\uparrow\rangle_{a1}|\downarrow\rangle_{b1}|\uparrow\rangle_{c1}\nonumber\\
&+&\gamma|\uparrow\rangle_{a1}|\uparrow\rangle_{b1}|\downarrow\rangle_{c1},\label{Wstate}
\end{eqnarray}
where $\alpha$, $\beta$ and $\gamma$ are the initial entanglement coefficients, and $|\alpha|^{2}+|\beta|^{2}+|\gamma|^{2}=1$. Here
$|\uparrow\rangle$ and $|\downarrow\rangle$ represent spin up and spin
down, respectively.
\begin{figure}[!h]
\begin{center}
\includegraphics[width=9cm,angle=0]{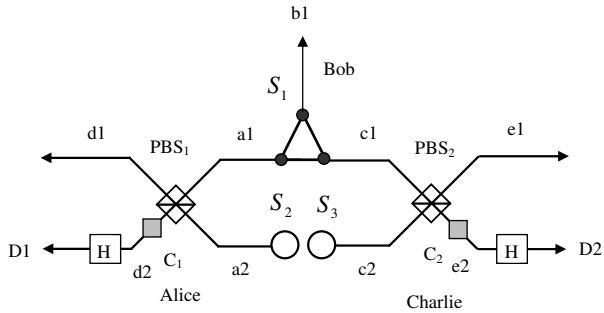}
\caption{A schematic drawing of our ECP with charge detectors.
$S_{1}$ is the less entanglement source and $S_{2}$ and $S_{3}$ are
the single electron sources. C$_{1}$ and C$_{2}$ are charge
detectors.  PBS$_{1}$ and PBS$_{2}$ transmit the spin up
$|\uparrow\rangle$ component and reflect the spin down
$|\downarrow\rangle$ component. a1, b1, c1, etc. are the different
spatial modes.}
\end{center}
\end{figure}
Then S$_{2}$ emits another single flying charge qubit and sends it to Alice with the form
\begin{eqnarray}
|\Phi\rangle_{a2}&=&\frac{\alpha}{\sqrt{|\alpha|^{2}+|\beta|^{2}}}|\uparrow\rangle_{a2}+\frac{\beta}{\sqrt{|\alpha|^{2}+|\beta|^{2}}}|\downarrow\rangle_{a2}.\label{auxiliary1}
\end{eqnarray}
The less-entangled W state combined with the flying qubit can be
described as
\begin{eqnarray}
|\Psi\rangle&=&|\Phi\rangle_{a1b1c1}\otimes|\Phi\rangle_{a2}=(\alpha|\downarrow\rangle_{a1}|\uparrow\rangle_{b1}|\uparrow\rangle_{c1}\nonumber\\
&+&\beta|\uparrow\rangle_{a1}|\downarrow\rangle_{b1}|\uparrow\rangle_{c1}
+\gamma|\uparrow\rangle_{a1}|\uparrow\rangle_{b1}|\downarrow\rangle_{c1})\nonumber\\
&\otimes&(\frac{\alpha}{\sqrt{|\alpha|^{2}+|\beta|^{2}}}|\uparrow\rangle_{a2}+\frac{\beta}{\sqrt{|\alpha|^{2}+|\beta|^{2}}}|\downarrow\rangle_{a2})\nonumber\\
&=&\frac{\alpha^{2}}{\sqrt{|\alpha|^{2}+|\beta|^{2}}}|\downarrow\rangle_{a1}|\uparrow\rangle_{a2}|\uparrow\rangle_{b1}|\uparrow\rangle_{c1}\nonumber\\
&+&\frac{\beta^{2}}{\sqrt{|\alpha|^{2}+|\beta|^{2}}}|\uparrow\rangle_{a1}|\downarrow\rangle_{a2}|\downarrow\rangle_{b1}|\uparrow\rangle_{c1}\nonumber\\
&+&\frac{\alpha\gamma}{\sqrt{|\alpha|^{2}+|\beta|^{2}}}|\uparrow\rangle_{a1}|\uparrow\rangle_{a2}|\uparrow\rangle_{b1}|\downarrow\rangle_{c1}\nonumber\\
&+&\frac{\beta\gamma}{\sqrt{|\alpha|^{2}+|\beta|^{2}}}|\uparrow\rangle_{a1}|\downarrow\rangle_{a2}|\uparrow\rangle_{b1}|\downarrow\rangle_{c1}\nonumber\\
&+&\frac{\alpha\beta}{\sqrt{|\alpha|^{2}+|\beta|^{2}}}|\downarrow\rangle_{a1}|\downarrow\rangle_{a2}|\uparrow\rangle_{b1}|\uparrow\rangle_{c1}\nonumber\\
&+&\frac{\alpha\beta}{\sqrt{|\alpha|^{2}+|\beta|^{2}}}|\uparrow\rangle_{a1}|\uparrow\rangle_{a2}|\downarrow\rangle_{b1}|\uparrow\rangle_{c1}.\label{combine1}
\end{eqnarray}
Alice lets the electrons in the spatial modes a1 and a2 pass through the PBS$_{1}$, which can
transmit the spin up $|\uparrow\rangle$ and reflect the spin down
$|\downarrow\rangle$, respectively. From Eq. (\ref{combine1}), after
the electrons passing through the charge detector $C_{1}$, one can
find that the item
$|\downarrow\rangle_{a1}|\uparrow\rangle_{a2}|\uparrow\rangle_{b1}|\uparrow\rangle_{c1}$
will lead $C_{1}$ get none electron. The items
$|\uparrow\rangle_{a1}|\downarrow\rangle_{a2}|\downarrow\rangle_{b1}|\uparrow\rangle_{c1}$
and
$|\uparrow\rangle_{a1}|\downarrow\rangle_{a2}|\uparrow\rangle_{b1}|\downarrow\rangle_{c1}$
will lead $C_{1}$ get both electrons. The items
$|\uparrow\rangle_{a1}|\uparrow\rangle_{a2}|\uparrow\rangle_{b1}|\downarrow\rangle_{c1}$,
$|\downarrow\rangle_{a1}|\downarrow\rangle_{a2}|\uparrow\rangle_{b1}|\uparrow\rangle_{c1}$
and
$|\uparrow\rangle_{a1}|\uparrow\rangle_{a2}|\downarrow\rangle_{b1}|\uparrow\rangle_{c1}$
will lead $C_{1}$ get only one electron. Then, if Alice chooses the
cases that the charge detector $C_{1}$ only obtains one electron,
 Eq. (\ref{combine1}) will become
\begin{eqnarray}
|\Psi\rangle'&=&\frac{\alpha\gamma}{\sqrt{|\alpha|^{2}+|\beta|^{2}}}|\uparrow\rangle_{d1}|\uparrow\rangle_{d2}|\uparrow\rangle_{b1}|\downarrow\rangle_{c1}\nonumber\\
&+&\frac{\alpha\beta}{\sqrt{|\alpha|^{2}+|\beta|^{2}}}|\downarrow\rangle_{d1}|\downarrow\rangle_{d2}|\uparrow\rangle_{b1}|\uparrow\rangle_{c1}\nonumber\\
&+&\frac{\alpha\beta}{\sqrt{|\alpha|^{2}+|\beta|^{2}}}|\uparrow\rangle_{d1}|\uparrow\rangle_{d2}|\downarrow\rangle_{b1}|\uparrow\rangle_{c1},\label{collpase1}
\end{eqnarray}
with the probability
\begin{eqnarray}
P_{1}=\frac{|\alpha|^{2}(|\gamma|^{2}+2|\beta|^{2})}{|\alpha|^{2}+|\beta|^{2}}.
\end{eqnarray}
$|\Psi\rangle'$ can be normalized as
\begin{eqnarray}
|\Psi\rangle''&=&\frac{\gamma}{\sqrt{|\gamma|^{2}+2|\beta|^{2}}}|\uparrow\rangle_{d1}|\uparrow\rangle_{d2}|\uparrow\rangle_{b1}|\downarrow\rangle_{c1}\nonumber\\
&+&\frac{\beta}{\sqrt{|\gamma|^{2}+2|\beta|^{2}}}|\downarrow\rangle_{d1}|\downarrow\rangle_{d2}|\uparrow\rangle_{b1}|\uparrow\rangle_{c1}\nonumber\\
&+&\frac{\beta}{\sqrt{|\gamma|^{2}+2|\beta|^{2}}}|\uparrow\rangle_{d1}|\uparrow\rangle_{d2}|\downarrow\rangle_{b1}|\uparrow\rangle_{c1}.\label{collpase2}
\end{eqnarray}
Then Alice takes the Hadamard operation (H) on the electron in mode d2, which can make
\begin{eqnarray}
|\uparrow\rangle\rightarrow\frac{1}{\sqrt{2}}(|\uparrow\rangle+|\downarrow\rangle),\nonumber\\
|\downarrow\rangle\rightarrow\frac{1}{\sqrt{2}}(|\uparrow\rangle-|\downarrow\rangle).
\end{eqnarray}
Finally, Alice measures the electron in the detector D$_{1}$ on the
$\{|\uparrow\rangle,|\downarrow\rangle\}$ basis. It can be found if the measurement
result is $|\uparrow\rangle$, they will get
\begin{eqnarray}
|\Phi_{1}\rangle_{d1b1c1}&=&\frac{\gamma}{\sqrt{|\gamma|^{2}+2|\beta|^{2}}}|\uparrow\rangle_{d1}|\uparrow\rangle_{b1}|\downarrow\rangle_{c1}\nonumber\\
&+&\frac{\beta}{\sqrt{|\gamma|^{2}+2|\beta|^{2}}}|\downarrow\rangle_{d1}|\uparrow\rangle_{b1}|\uparrow\rangle_{c1}\nonumber\\
&+&\frac{\beta}{\sqrt{|\gamma|^{2}+2|\beta|^{2}}}|\uparrow\rangle_{d1}|\downarrow\rangle_{b1}|\uparrow\rangle_{c1},\label{concentrated1}
\end{eqnarray}
while if the measurement result is $|\downarrow\rangle$, they will get
\begin{eqnarray}
|\Phi_{2}\rangle_{d1b1c1}&=&\frac{\gamma}{\sqrt{|\gamma|^{2}+2|\beta|^{2}}}|\uparrow\rangle_{d1}|\uparrow\rangle_{b1}|\downarrow\rangle_{c1}\nonumber\\
&-&\frac{\beta}{\sqrt{|\gamma|^{2}+2|\beta|^{2}}}|\downarrow\rangle_{d1}|\uparrow\rangle_{b1}|\uparrow\rangle_{c1}\nonumber\\
&+&\frac{\beta}{\sqrt{|\gamma|^{2}+2|\beta|^{2}}}|\uparrow\rangle_{d1}|\downarrow\rangle_{b1}|\uparrow\rangle_{c1}.\label{concentrated2}
\end{eqnarray}
If they got $|\Phi_{2}\rangle_{d1b1c1}$, one of the three parties, say Alice, Bob or Charlie only should
perform a local phase rotation operation on her or his electron, they can get $|\Phi_{1}\rangle_{d1b1c1}$.

After Alice gets the $|\Phi_{1}\rangle$ successfully, we start the second concentration step. The electron source S$_{3}$ emits another single electron and sends it to Charlie in the spatial mode $c2$. The single electron is with the form
\begin{eqnarray}
|\Phi\rangle_{c2}&=&\frac{\beta}{\sqrt{|\gamma|^{2}+|\beta|^{2}}}|\downarrow\rangle_{c2}+\frac{\gamma}{\sqrt{|\gamma|^{2}+|\beta|^{2}}}|\uparrow\rangle_{c2}.\label{auxiliary3}
\end{eqnarray}

After Charlie receives the single mobile electron, the whole system
can be described as:
\begin{eqnarray}
&&|\Phi_{1}\rangle_{d1b1c1}\otimes|\Phi\rangle_{c2}\nonumber\\
&=&\frac{\beta\gamma}{\sqrt{|\gamma|^{2}+2|\beta|^{2}}\sqrt{|\gamma|^{2}+|\beta|^{2}}}|\uparrow\rangle_{d1}|\uparrow\rangle_{b1}|\downarrow\rangle_{c1}|\downarrow\rangle_{c2}\nonumber\\
&+&\frac{\gamma^{2}}{\sqrt{|\gamma|^{2}+2|\beta|^{2}}\sqrt{|\gamma|^{2}+|\beta|^{2}}}|\uparrow\rangle_{d1}|\uparrow\rangle_{b1}|\downarrow\rangle_{c1}|\uparrow\rangle_{c2}\nonumber\\
&+&\frac{\beta^{2}}{\sqrt{|\gamma|^{2}+2|\beta|^{2}}\sqrt{|\gamma|^{2}+|\beta|^{2}}}|\downarrow\rangle_{d1}|\uparrow\rangle_{b1}|\uparrow\rangle_{c1}|\downarrow\rangle_{c2}\nonumber\\
&+&\frac{\beta\gamma}{\sqrt{|\gamma|^{2}+2|\beta|^{2}}\sqrt{|\gamma|^{2}+|\beta|^{2}}}|\downarrow\rangle_{d1}|\uparrow\rangle_{b1}|\uparrow\rangle_{c1}|\uparrow\rangle_{c2}\nonumber\\
&+&\frac{\beta^{2}}{\sqrt{|\gamma|^{2}+2|\beta|^{2}}\sqrt{|\gamma|^{2}+|\beta|^{2}}}|\uparrow\rangle_{d1}|\downarrow\rangle_{b1}|\uparrow\rangle_{c1}|\downarrow\rangle_{c2}\nonumber\\
&+&\frac{\beta\gamma}{\sqrt{|\gamma|^{2}+2|\beta|^{2}}\sqrt{|\gamma|^{2}+|\beta|^{2}}}|\uparrow\rangle_{d1}|\downarrow\rangle_{b1}|\uparrow\rangle_{c1}|\uparrow\rangle_{c2}.\nonumber\\\label{collpase3}
\end{eqnarray}

Then, Charlie makes the electrons in the spatial modes c1 and c2 pass through the PBS$_{2}$. It is interesting to find that items
$|\uparrow\rangle_{d1}|\uparrow\rangle_{b1}|\downarrow\rangle_{c1}|\uparrow\rangle_{c2}$
will lead both the two electrons in Charlie's location in the
spatial mode e1, which makes the charge detector C$_{2}$ cannot detect any
electrons. The items $|\downarrow\rangle_{d1}|\uparrow\rangle_{b1}|\uparrow\rangle_{c1}|\downarrow\rangle_{c2}$
and $|\uparrow\rangle_{d1}|\downarrow\rangle_{b1}|\uparrow\rangle_{c1}|\downarrow\rangle_{c2}$
will lead both the two electrons in the $e2$ mode, which makes the charge
detector C$_{2}$ detect two electrons. All the three items $|\uparrow\rangle_{d1}|\uparrow\rangle_{b1}|\downarrow\rangle_{c1}|\downarrow\rangle_{c2}$,
$|\downarrow\rangle_{d1}|\uparrow\rangle_{b1}|\uparrow\rangle_{c1}|\uparrow\rangle_{c2}$
and $|\uparrow\rangle_{d1}|\downarrow\rangle_{b1}|\uparrow\rangle_{c1}|\uparrow\rangle_{c2}$
make the spatial mode $e2$ contains only one electron, which makes the detector C$_{2}$ detect exactly one electron. Then Charlie selects the items which makes
C$_{2}$ detect only one electron, then the whole state will essentially collapse to the four-electron maximally entangled W state:
\begin{eqnarray}
|\Psi'''\rangle&=&\frac{1}{\sqrt{3}}(|\uparrow\rangle_{d1}|\uparrow\rangle_{b1}|\downarrow\rangle_{e1}|\downarrow\rangle_{e2}\nonumber\\
&+&|\downarrow\rangle_{d1}|\uparrow\rangle_{b1}|\uparrow\rangle_{e1}|\uparrow\rangle_{e2}+|\uparrow\rangle_{d1}|\downarrow\rangle_{b1}|\uparrow\rangle_{e1}|\uparrow\rangle_{e2}),\nonumber\\
\end{eqnarray}
with the probability of
\begin{eqnarray}
P_{2}=\frac{3|\beta|^{2}|\gamma|^{2}}{(|\gamma|^{2}+|\beta|^{2})(|\gamma|^{2}+2|\beta|^{2})},
\end{eqnarray}
where the subscription 2 means in the second concentration step.

Then, similar to the first step, Charlie performs the Hadamard operation on the electron in the e2 mode. After the Hadamard operation, he detects the electron in the e2 mode on the $\{|\uparrow\rangle,|\downarrow\rangle\}$ basis by the detector D$_{2}$.
If the measurement result is $|\uparrow\rangle$, they will get
\begin{eqnarray}
|\Phi_{1}\rangle_{d1b1e1}&=&\frac{1}{\sqrt{3}}(|\uparrow\rangle_{d1}|\uparrow\rangle_{b1}|\downarrow\rangle_{e1}\nonumber\\
&+&|\downarrow\rangle_{d1}|\uparrow\rangle_{b1}|\uparrow\rangle_{e1}+|\uparrow\rangle_{d1}|\downarrow\rangle_{b1}|\uparrow\rangle_{e1}).\nonumber\\\label{concentrated3}
\end{eqnarray}
 Otherwise, if the
measurement result is $|\downarrow\rangle$, they will get
\begin{eqnarray}
|\Phi_{2}\rangle_{d1b1e1}&=&\frac{1}{\sqrt{3}}(-|\uparrow\rangle_{d1}|\uparrow\rangle_{b1}|\downarrow\rangle_{e1}\nonumber\\
&+&|\downarrow\rangle_{d1}|\uparrow\rangle_{b1}|\uparrow\rangle_{e1}+|\uparrow\rangle_{d1}|\downarrow\rangle_{b1}|\uparrow\rangle_{e1}).\nonumber\\\label{concentrated4}
\end{eqnarray}
 In this case, one of the three parties, say Alice, Bob or Charlie should perform a local
phase rotation operation on her or his electron, then they can get $|\Phi_{1}\rangle_{d1b1e1}$. So far, the whole ECP is completed. We can
calculate its total success probability, which equals the product of the probability in each concentration step as
\begin{eqnarray}
P=P_{1}P_{2}&=&\frac{(|\gamma|^{2}+2|\beta|^{2})}{|\alpha|^{2}+|\beta|^{2}}
\cdot\frac{3|\beta|^{2}|\gamma|^{2}}{(|\gamma|^{2}+|\beta|^{2})(|\gamma|^{2}+2|\beta|^{2})}\nonumber\\
&=&\frac{3|\alpha|^{2}|\beta|^{2}|\gamma|^{2}}{(|\alpha|^{2}+|\beta|^{2})(|\gamma|^{2}+|\beta|^{2})}.
\end{eqnarray}
\section{The second ECP with complete parity check gate}

By far, we have described our ECP with PBSs and charge detectors. In
fact, in the ECP, PBS combined with charge detectors acts the role of
parity checking. During the above description, Alice and Charlie
essentially pick up the even parity states
$|\uparrow\rangle|\uparrow\rangle$ and
$|\downarrow\rangle|\downarrow\rangle$, but discard the odd parity
states $|\uparrow\rangle|\downarrow\rangle$ and
$|\downarrow\rangle|\uparrow\rangle$, for after the PBS, the
$|\uparrow\rangle|\downarrow\rangle$ and
$|\downarrow\rangle|\uparrow\rangle$ will lead the two electrons both in
the same spatial mode. We take Eq. (\ref{combine1}) as an example.
One can find that the item
$|\downarrow\rangle_{a1}|\uparrow\rangle_{a2}|\uparrow\rangle_{b1}|\uparrow\rangle_{c1}$
will lead $C_{1}$ get none electron, and the items
$|\uparrow\rangle_{a1}|\downarrow\rangle_{a2}|\downarrow\rangle_{b1}|\uparrow\rangle_{c1}$
and
$|\uparrow\rangle_{a1}|\downarrow\rangle_{a2}|\uparrow\rangle_{b1}|\downarrow\rangle_{c1}$
will lead $C_{1}$ get both electrons. Therefore, as only one PBS
combined with charge detector can only pick up the even parity states but has to discard the
odd states, it cannot make a complete parity check.
\begin{figure}[!h]
\begin{center}
\includegraphics[width=8cm,angle=0]{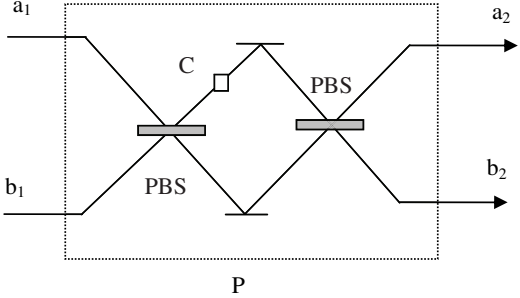}
\caption{A schematic drawing of our complete parity check gate (P).
It is also shown in Ref. \cite{Beenakker}. It can completely
distinguish the even parity states
$|\uparrow\rangle|\uparrow\rangle$ and
$|\downarrow\rangle|\downarrow\rangle$ from the odd parity states
$|\uparrow\rangle|\downarrow\rangle$ and
$|\downarrow\rangle|\uparrow\rangle$.}
\end{center}
\end{figure}

In this section, we will describe another ECP with the complete parity check gate. Before we start to explain this ECP, we briefly explain the complete parity check, as shown in Fig. 2. In fact, this parity check has been studied for
several years to construct the controlled-not(CNOT) gate
\cite{Beenakker}, and perform the  entanglement concentration and
purification \cite{shengpla}. From Fig. 2, We suppose that two
electrons $|\varphi_{1}\rangle=\alpha_{1}|\uparrow\rangle_{a_{1}}+\beta_{1}|\downarrow\rangle_{a_{1}}$
and $|\varphi_{2}\rangle=\alpha_{2}|\uparrow\rangle_{b_{1}}+\beta_{2}|\downarrow\rangle_{b_{1}}$
 enter the complete parity check gate from the spatial modes
 a1 and  b1, respectively. The whole state can be described as
\begin{eqnarray}
|\varphi_{1}\rangle\otimes|\varphi_{2}\rangle&=&(\alpha_{1}|\uparrow\rangle_{a_{1}}+\beta_{1}|\downarrow\rangle_{a_{1}})
\otimes(\alpha_{2}|\uparrow\rangle_{b_{1}}+\beta_{2}|\downarrow\rangle_{b_{1}})\nonumber\\
&=&\alpha_{1}\alpha_{2}|\uparrow\rangle_{a_{1}}|\uparrow\rangle_{b_{1}}+\beta_{1}\beta_{2}|\downarrow\rangle_{a_{1}}|\downarrow\rangle_{b_{1}}\nonumber\\
&+&\alpha_{1}\beta_{2}|\uparrow\rangle_{a_{1}}|\downarrow\rangle_{b_{1}}+\beta_{1}\alpha_{2}|\downarrow\rangle_{a_{1}}|\uparrow\rangle_{b_{1}}.
\end{eqnarray}
It is obvious that items
$\alpha_{1}\alpha_{2}|\uparrow\rangle_{a_{1}}|\uparrow\rangle_{b_{1}}+\beta_{1}\beta_{2}|\downarrow\rangle_{a_{1}}|\downarrow\rangle_{b_{1}}$
will lead $C=1$. After the complete parity check gate, they will finally convert to
$\alpha_{1}\alpha_{2}|\uparrow\rangle_{a_{2}}|\uparrow\rangle_{b_{2}}+\beta_{1}\beta_{2}|\downarrow\rangle_{a_{2}}|\downarrow\rangle_{b_{2}}$.
The items
$\alpha_{1}\beta_{2}|\uparrow\rangle_{a_{1}}|\downarrow\rangle_{b_{1}}+\beta_{1}\alpha_{2}|\downarrow\rangle_{a_{1}}|\uparrow\rangle_{b_{1}}$
will lead $C=0$. After the complete parity check gate, they will become
$\alpha_{1}\beta_{2}|\uparrow\rangle_{a_{2}}|\downarrow\rangle_{b_{2}}+\beta_{1}\alpha_{2}|\downarrow\rangle_{a_{2}}|\uparrow\rangle_{b_{2}}$.
That is to say, the gate can completely distinguish the even parity state
$\alpha_{1}\alpha_{2}|\uparrow\rangle_{a_{1}}|\uparrow\rangle_{b_{1}}+\beta_{1}\beta_{2}|\downarrow\rangle_{a_{1}}|\downarrow\rangle_{b_{1}}$
from the odd parity state
$\alpha_{1}\beta_{2}|\uparrow\rangle_{a_{1}}|\downarrow\rangle_{b_{1}}+\beta_{1}\alpha_{2}|\downarrow\rangle_{a_{1}}|\uparrow\rangle_{b_{1}}$.
\begin{figure}[!h]
\begin{center}
\includegraphics[width=7cm,angle=0]{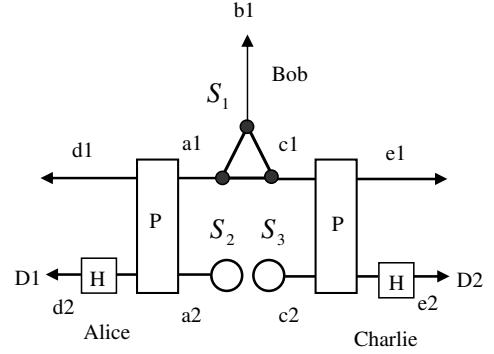}
\caption{A schematic drawing of our ECP with P gates. We adopt the P
gates shown in Fig. 2 to replace the PBSs to reconstruct this ECP
and make it have higher success probability.}
\end{center}
\end{figure}

Now we use the complete parity check gate to substitute the PBS to
reperform the ECP. We denote it as P gate in Fig. 2.  From Eq.
(\ref{combine1}), Alice makes the electrons in a1 and a2 modes pass through the P gate. After the P gate, if the charge
detector shows $C=1$, they will get the same state in Eq.
(\ref{collpase1}), which can be finally converted to Eq. (\ref{concentrated1}) or Eq.
(\ref{concentrated2}) following the same step described above. On the other hand, if the charge detector
shows $C=0$, they will get a four-electron less-entangled state as
\begin{eqnarray}
|\Psi_{1}\rangle'_{d1d2b1c1}&=&\frac{\alpha^{2}}{\sqrt{|\alpha|^{2}+|\beta|^{2}}}|\downarrow\rangle_{d1}|\uparrow\rangle_{d2}|\uparrow\rangle_{b1}|\uparrow\rangle_{c1}\nonumber\\
&+&\frac{\beta^{2}}{\sqrt{|\alpha|^{2}+|\beta|^{2}}}|\uparrow\rangle_{d1}|\downarrow\rangle_{d2}|\downarrow\rangle_{b1}|\uparrow\rangle_{c1}\nonumber\\
&+&\frac{\beta\gamma}{\sqrt{|\alpha|^{2}+|\beta|^{2}}}|\uparrow\rangle_{d1}|\downarrow\rangle_{d2}|\uparrow\rangle_{b1}|\downarrow\rangle_{c1}.\label{less0}\nonumber\\
\end{eqnarray}
Then, Alice measures the electron in $d2$ mode on the
$\{|\uparrow\rangle,|\downarrow\rangle\}$ basis. After the Hadamard
operation, Eq. (\ref{less0}) will convert to another less-entangled W state of the form
\begin{eqnarray}
|\Psi^{\pm}_{1}\rangle'_{a1b1c1}&=&\alpha'|\downarrow\rangle_{a1}|\uparrow\rangle_{b1}|\uparrow\rangle_{c1}
\pm\beta'|\uparrow\rangle_{a1}|\downarrow\rangle_{b1}|\uparrow\rangle_{c1}\nonumber\\
&\pm&\gamma'|\uparrow\rangle_{a1}|\uparrow\rangle_{b1}|\downarrow\rangle_{c1},\label{less1}
\end{eqnarray}
with
\begin{eqnarray}
\alpha'=\frac{\alpha^{4}}{\sqrt{|\alpha|^{4}+|\beta|^{4}+|\beta|^{2}|\gamma|^{2}}},\nonumber\\
\beta'=\frac{\beta^{4}}{\sqrt{|\alpha|^{4}+|\beta|^{4}+|\beta|^{2}|\gamma|^{2}}},\nonumber\\
\gamma'=\frac{\beta^{2}\gamma^{2}}{\sqrt{|\alpha|^{4}+|\beta|^{4}+|\beta|^{2}|\gamma|^{2}}}.
\end{eqnarray}
'+' or '-' depends on the measurement results. If the measurement result of
D1 is $|\uparrow\rangle$, it is '+', otherwise, it is '-'. The state
$|\Psi^{\pm}_{1}\rangle'_{a1b1c1}$ can be reconcentrated with
the same principle. Alice only needs to prepare another single
electron state from $S_{2}$ of the form
\begin{eqnarray}
|\Phi\rangle'_{a2}&=&\frac{\alpha'}{\sqrt{|\alpha'|^{2}+|\beta'|^{2}}}|\uparrow\rangle_{a2}
+\frac{\beta'}{\sqrt{|\alpha'|^{2}+|\beta'|^{2}}}|\downarrow\rangle_{a2}.\nonumber\\
\end{eqnarray}
 After making the electrons in the a1 and a2 modes pass through the P gate, if
the charge detector shows $C=1$, the combination
$|\Phi\rangle'_{a2}\otimes|\Psi^{+}_{1}\rangle'_{a1b1c1}$ will
become
\begin{eqnarray}
|\Psi_{1}\rangle'&=&\frac{\alpha'\gamma'}{\sqrt{|\alpha'|^{2}+|\beta'|^{2}}}|\uparrow\rangle_{d1}|\uparrow\rangle_{d2}|\uparrow\rangle_{b1}|\downarrow\rangle_{c1}\nonumber\\
&+&\frac{\alpha'\beta'}{\sqrt{|\alpha'|^{2}+|\beta'|^{2}}}|\downarrow\rangle_{d1}|\downarrow\rangle_{d2}|\uparrow\rangle_{b1}|\uparrow\rangle_{c1}\nonumber\\
&+&\frac{\alpha'\beta'}{\sqrt{|\alpha'|^{2}+|\beta'|^{2}}}|\uparrow\rangle_{d1}|\uparrow\rangle_{d2}|\downarrow\rangle_{b1}|\uparrow\rangle_{c1}.
\end{eqnarray}
Otherwise, if $C=0$, they will ultimately get another less-entangled W state as
\begin{eqnarray}
|\Psi^{\pm}_{1}\rangle''_{a1b1c1}&=&\alpha''|\downarrow\rangle_{a1}|\uparrow\rangle_{b1}|\uparrow\rangle_{c1}
\pm\beta''|\uparrow\rangle_{a1}|\downarrow\rangle_{b1}|\uparrow\rangle_{c1}\nonumber\\
&\pm&\gamma''|\uparrow\rangle_{a1}|\uparrow\rangle_{b1}|\downarrow\rangle_{c1},\label{less1}
\end{eqnarray}
with
\begin{eqnarray}
\alpha''=\frac{\alpha'^{4}}{\sqrt{|\alpha'|^{4}+|\beta'|^{4}+|\beta'|^{2}|\gamma'|^{2}}},\nonumber\\
\beta''=\frac{\beta^{4}}{\sqrt{|\alpha'|^{4}+|\beta'|^{4}+|\beta'|^{2}|\gamma'|^{2}}},\nonumber\\
\gamma''=\frac{\beta'^{2}\gamma'^{2}}{\sqrt{|\alpha'|^{4}+|\beta'|^{4}+|\beta'|^{2}|\gamma'|^{2}}}.
\end{eqnarray}

In this way, they can repeat the whole protocol to get a higher
success probability.
We can calculate the success probability in the second round as
\begin{eqnarray}
P_{2}^{1}&=&\frac{|\alpha|^{4}(|\beta|^{2}|\gamma|^{2}+2|\beta|^{4})}{(|\alpha|^{4}+|\beta|^{4})(|\alpha|^{2}+|\beta|^{2})}.\nonumber\\
\end{eqnarray}
We can also calculate the success probability in the third and other round as
\begin{eqnarray}
P_{3}^{1}&=&\frac{|\alpha|^{8}(|\beta|^{6}|\gamma|^{2}+2|\beta|^{8})}{(|\alpha|^{8}+|\beta|^{8})(|\alpha|^{4}+|\beta|^{4})(|\alpha|^{2}+|\beta|^{2})},\nonumber\\
&\cdots&\nonumber\\
P_{N}^{1}&=&\frac{|\alpha|^{2^{N}}(|\beta|^{2^{N}-2}|\gamma|^{2}+2|\beta|^{2^{N}})}{(|\alpha|^{2^{N}}+|\beta|^{2^{N}})(|\alpha|^{2^{N-1}}+|\beta|^{2^{N-1}})\cdots(|\alpha|^{2}+|\beta|^{2})}.\nonumber\\
\end{eqnarray}
Here the superscription $"1"$ means the first step performed by Alice. The subscription
$"1","2","3",\cdots "N"$ is the iteration number. Then the total success probability of the first concentration step equals the sum of the probability in each round, which can be written as 
\begin{eqnarray}
P^{1}=\sum_{N=1}^{\infty}P^{1}_{N}.
\end{eqnarray}

The same principle can also be used in Charlie's location. For
example, after passing
through the P gate, if $C=1$, $|\Phi_{1}\rangle_{d1b1c1}\otimes|\Phi\rangle_{c2}$ will
become the maximally entangled W state. Otherwise, if $C=0$, they will get
\begin{eqnarray}
&&|\Psi_{2}\rangle'_{d1b1c1e2}\nonumber\\
&=&\frac{\gamma^{2}}{\sqrt{|\gamma|^{2}+2|\beta|^{2}}\sqrt{|\gamma|^{2}+|\beta|^{2}}}|\uparrow\rangle_{d1}|\uparrow\rangle_{b1}|\downarrow\rangle_{c1}|\uparrow\rangle_{e2}\nonumber\\
&+&\frac{\beta^{2}}{\sqrt{|\gamma|^{2}+2|\beta|^{2}}\sqrt{|\gamma|^{2}+|\beta|^{2}}}|\downarrow\rangle_{d1}|\uparrow\rangle_{b1}|\uparrow\rangle_{c1}|\downarrow\rangle_{e2}\nonumber\\
&+&\frac{\beta^{2}}{\sqrt{|\gamma|^{2}+2|\beta|^{2}}\sqrt{|\gamma|^{2}+|\beta|^{2}}}|\uparrow\rangle_{d1}|\downarrow\rangle_{b1}|\uparrow\rangle_{c1}|\downarrow\rangle_{e2}.\nonumber\\\label{less20}
\end{eqnarray}
After performing the Hadamard operation and measuring the electron
in e2 mode, it will become
\begin{eqnarray}
|\Psi^{\pm}_{2}\rangle'_{d1b1c1}&=&\gamma''|\uparrow\rangle_{d1}|\uparrow\rangle_{b1}\downarrow\rangle_{c1}\pm\beta''|\downarrow\rangle_{d1}|\uparrow\rangle_{b1}|\uparrow\rangle_{c1}\nonumber\\
&\pm&\beta''|\uparrow\rangle_{d1}|\downarrow\rangle_{b1}|\uparrow\rangle_{c1},\label{less2}
\end{eqnarray}
with
\begin{eqnarray}
\gamma''=\frac{\gamma^{2}}{\sqrt{|\gamma|^{4}+2|\beta|^{4}}},\nonumber\\
\beta''=\frac{\beta^{2}}{\sqrt{|\gamma|^{4}+2|\beta|^{4}}}.
\end{eqnarray}
'+' or '-' depends on the measurement result. If the measurement
result in D2 is $|\uparrow\rangle$, it is '+', otherwise, it is '-'.
If they get $|\Psi^{+}_{2}\rangle'_{d1b1c1}$, they only need to
prepare another single electron of the form
\begin{eqnarray}
|\Phi'\rangle_{c2}=\frac{\beta^{2}}{\sqrt{|\gamma|^{4}+|\beta|^{4}}}|\downarrow\rangle_{c2}+\frac{\gamma^{2}}{\sqrt{|\gamma|^{4}+|\beta|^{4}}}|\uparrow\rangle_{c2}.
\end{eqnarray}
Following the same principle described above, the combination state
$|\Psi^{+}_{2}\rangle'_{d1b1c1}\otimes|\Phi'\rangle_{c2}$ can also
ultimately become the maximally entangled W state with some probability. In this way, we have proved that the second concentration step can also be repeated to get a higher success probability.We can calculate the success probability in each concentration round as
\begin{eqnarray}
P_{1}^{2}&=&\frac{3|\beta|^{2}|\gamma|^{2}}{(|\gamma|^{2}+|\beta|^{2})(|\gamma|^{2}+2|\beta|^{2}|)},\nonumber\\
P_{2}^{2}&=&\frac{3|\beta|^{4}|\gamma|^{4}}{(|\gamma|^{2}+2|\beta^{2}|)(|\gamma|^{4}+|\beta|^{4})(|\gamma|^{2}+|\beta|^{2})},\nonumber\\
P_{3}^{2}&=&\frac{3|\beta|^{8}|\gamma|^{8}}{(|\gamma|^{2}+2|\beta|^{2})(|\gamma|^{8}+|\beta|^{8})(|\gamma|^{4}+|\beta|^{4})(|\gamma|^{2}+|\beta|^{2})},\nonumber\\
&\cdots&\nonumber\\
P_{M}^{2}
&=&\frac{3|\beta|^{2^{M}}|\gamma|^{2^{M}}}{(|\gamma|^{2^{M}}+|\beta|^{2^{M}})(|\gamma|^{2^{M-1}}+|\beta|^{2^{M-1}})\cdots(|\gamma|^{2}+|\beta|^{2})}\nonumber\\
&\cdot&\frac{1}{(|\gamma|^{2}+2|\beta|^{2})}.
\end{eqnarray}
Here the superscription $"2"$ means the second step performed by Charlie. The subscription
$"1","2","3",\cdots "M"$ is the iteration number. The total success probability of the second concentration step equals the sum of the probability in each round, which can be written as
\begin{eqnarray}
P^{2}&=&\sum_{M=1}^{\infty}P^{2}_{M}
\end{eqnarray}

Finally, the total success probability of the whole ECP equals the product of the probability in each two concentration step as
\begin{eqnarray}
P_{T}=\sum_{N=1}^{\infty}P^{1}_{N}\sum_{M=1}^{\infty}P^{2}_{M}.\label{total}
\end{eqnarray}

\begin{figure}[!h]
\begin{center}
\includegraphics[width=8cm,angle=0]{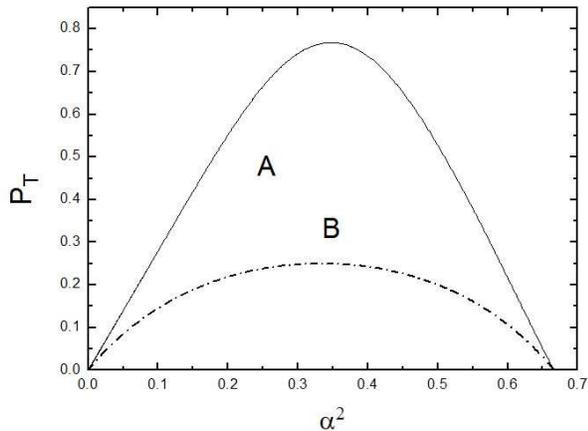}
\caption{The total success probability $P_{T}$ for getting a maximally
entangled W state is altered with the initial coefficient $\alpha^{2}$.
Here, we choose $\beta^{2}=\frac{1}{3}$,
$\alpha^{2}\in(0,\frac{2}{3})$. Curve A represents the total success probability of the second ECP, while
Curve B represents that of the first ECP. In the curve A, for numerical simulation, we choose
$N=M=3$ for the approximation.}
\end{center}
\end{figure}

The total success probability of the two ECPs as a function of the $\alpha^{2}$ is shown in Fig. 4.
 We choose $\beta^{2}=\frac{1}{3}$, and change
$\alpha^{2}\in(0,\frac{2}{3})$. For the second ECP, we choose
$N=M=3$ for a good numerical simulation. It is shown that when $\alpha^{2}\in(0,\frac{1}{3})$
both success probability monotonic increase
with $\alpha^{2}$. They both have a
maximally value when $\alpha^{2}=\frac{1}{3}$. Moreover, it is obvious that by repeating the second ECP, the total success probability
can be increased largely.

\section{discussion and summary}

By far, we have fully described our two ECPs for three-electron
 less-entangled W states. The two ECPs only need one pair of less-entangled W state, while other ECPs require two pairs of such states. It makes our two ECPs more economic than others. In the first ECP, we adopt the PBS combined
 with the charge detector to pick up the even parity states. However, the odd parity states have to be discarded. Therefore, it is not an optimal ECP. The second ECP is an improved protocol. In the second ECP, with the help of two
 PBSs and charge detectors, both the even parity states and the odd parity
 states can be remained. By repeating the ECP, the odd parity states can be reconcentrated to the maximally entangled states with some probability. So the whole protocol can reach a
 higher success probability than the first protocol. Certainly, in both ECPs,
 we should know the exact coefficients $\alpha$, $\beta$ and $\gamma$ in advance to prepare
 the auxiliary single electron state. According to Ref. \cite{dengpra}, in a practical concentration, the exact initial coefficients can be obtained by measuring enough amount of initial less-entangled samples\cite{dengpra}.

 Charge detection has played a prominent role in constructing the
 parity check gate. In
 fact, in Refs. \cite{Yamamoto1,zhao1}, the optical PBS can also act as the role of the
 parity check gate. However, in their protocols, after the photons
 passing through the PBS, one has to use the sophisticated
 single-photon detectors to ensure both spatial modes exactly
 contain only one photon. Moreover, the electrons will be destroyed if it is detected by the detector, which is so called the post-selection principle.
 Interestingly, in our first protocol, the charge detector essentially
 acts the similar role as the sophisticated single-photon detector. If its measurement result is $C=1$, the two spatial modes will exactly contain one electron. Moreover, the charge detection can not destroy the electron, that is to say, after the charge detection, the electrons can be remained for other application. Under present experimental conditions, the charge
 detection can be realized by means of the point contacts in a
 two-dimensional electron gas \cite{field}. Elzerman \emph{et al.} once showed that the
 current achievable time resolution for the charge detection is $\mu s$ \cite{Elzerman}.
Trauzettel \emph{et al.} put forward a protocol for realizing
such charge parity meter by using two double quantum dots along side a quantum point contact \cite{trauzettel}. Mao \emph{et al.}
proposed a more universal method to realize such device \cite{mao}. Another important element of these ECPs is the PBS in spin.
Ref. \cite{elecPBS} described a such device with the help of the beam splitters and the  electric field. The total efficiency of the PBS
can reach 100\% in principle. In 2004, based on the PBS in spin, the electrical quantum computation protocol with mobile spin qubits was proposed \cite{elecPBS2}.

In summary, we have proposed two ECPs for three-electron less-entangled W
state with the help of the charge detection. Both the two protocols only require one pair of less-entangled W state, which makes them more economical. In the first ECP, by picking up the even parity states, but discarding the odd parity states, we can ultimately obtain the maximally entangled W state with some success probability. The second ECP is an improved one, for it can be used repeatedly to reconcentrate the odd parity states and obtain a higher success probability. These features will make
these ECPs have a practical application in current solid quantum
computation and communication.

 \section*{ACKNOWLEDGEMENTS}
This work is supported by the National Natural Science Foundation of
China under Grant No. 11104159, and 61201164, Open Research
Fund Program of the State Key Laboratory of
Low-Dimensional Quantum Physics Scientific, Tsinghua University,
Open Research Fund Program of National Laboratory of Solid State Microstructures under Grant No. M25020 and M25022,
 A Project
Funded by the Priority Academic Program Development of Jiangsu
Higher Education Institutions.

\end{document}